\documentclass[prd,twocolumn,nofootinbib,aps,tightenlines,preprintnumbers,notitlepage,longbibliography,superscriptaddress]{revtex4-1}
\usepackage{graphicx}
\usepackage{amsmath}
\usepackage{amsfonts}
\usepackage[colorlinks=true,citecolor=blue,urlcolor=cyan,linkcolor=black]{hyperref}
\usepackage{graphicx}
\usepackage{dcolumn}
\usepackage{bm}
\usepackage[usenames, dvipsnames]{color}
\usepackage[normalem]{ulem}
\usepackage{xcolor}
\usepackage{subfigure}

\usepackage{footnote}

\graphicspath{ {figures/} }

\usepackage{epstopdf}
\newcommand{\be}{\begin{eqnarray}}
\newcommand{\non}{\nonumber \\}
\newcommand{\ee}{\end{eqnarray}}

\usepackage[export]{adjustbox}

\usepackage{float}

\usepackage{suffix}
\usepackage{mathtools}

\DeclarePairedDelimiterX\MeijerM[3]{\lparen}{\rparen}%
{\begin{smallmatrix}#1 \\ #2\end{smallmatrix}\delimsize\vert\,#3}

\newcommand\MeijerG[8][]{%
  G^{\,#2,#3}_{#4,#5}\MeijerM[#1]{#6}{#7}{#8}}

\WithSuffix\newcommand\MeijerG*[7]{%
  G^{\,#1,#2}_{#3,#4}\MeijerM*{#5}{#6}{#7}}

\usepackage{hyperref}
\hypersetup{
    bookmarks=true,         % show bookmarks bar?
    unicode=false,          % non-Latin characters in Acrobatâs bookmarks
    pdftoolbar=true,        % show Acrobatâs toolbar?
    pdfmenubar=true,        % show Acrobatâs menu?
    pdffitwindow=false,     % window fit to page when opened
    pdfstartview={FitH},    % fits the width of the page to the window
    pdftitle={My title},    % title
    pdfauthor={Author},     % author
    pdfsubject={Subject},   % subject of the document
    pdfcreator={Creator},   % creator of the document
    pdfproducer={Producer}, % producer of the document
    pdfkeywords={keyword1, key2, key3}, % list of keywords
    pdfnewwindow=true,      % links in new PDF window
    colorlinks=true,       % false: boxed links; true: colored links
    linkcolor=black,          % color of internal links (change box color with linkbordercolor)
    citecolor=black,        % color of links to bibliography
    filecolor=magenta,      % color of file links
    urlcolor=cyan           % color of external links
}

\newcommand{\aap}{Astron. Astrophys.}

\newcommand{\aj}{Astron. J.}

\newcommand{\mnras}{Mon. Not. R. Astron. Soc.}

\newcommand{\physrep}{Phys. Rep.}

\newcommand{\dd}{{\rm d}}

\begin{document}

% \title{Probing compensated isocurvature fluctuations in the cosmic-dawn with the 21cm power-spectrum}

\title{Probing Compensated Isocurvature  with the 21-cm Signal during Cosmic Dawn}

\newcommand{\imperial}{Department of Physics, Imperial College London, Blackett Laboratory, Prince Consort Road, London SW7 2AZ, UK}

\newcommand{\jhu}{Department of Physics \& Astronomy, Johns Hopkins University, Baltimore, MD 21218, USA}

\newcommand{\tifr}{Department of Theoretical Physics, Tata Institute of Fundamental Research, Dr Homi Bhabha Road, Navy Nagar, Colaba, Mumbai-400005, India}

\newcommand{\harvard}{Harvard-Smithsonian Center for Astrophysics, 60 Garden St., Cambridge, MA 02138, USA}

\author{Selim~C.~Hotinli}
\email{selim.hotinli14@imperial.ac.uk}
\affiliation{\jhu}

\author{Thomas~Binnie}
\email{t.binnie16@imperial.ac.uk}
\affiliation{\imperial}

\author{Julian~B.~Mu$\tilde{\rm n}$oz}
\email{julianmunoz@cfa.harvard.edu}
\affiliation{\harvard}

\author{Bikash~R.~Dinda}
\email{bikashd18@gmail.com}
\affiliation{\tifr}

\author{Marc~Kamionkowski}
\email{kamion@jhu.edu}
\affiliation{\jhu}

\date{\today}

\begin{abstract}

Upcoming measurements of the 21-cm line of neutral hydrogen will open a new observational window into the early stages of structure growth, providing a unique opportunity for probing large-scale cosmological signatures using the small-scale signals from the first stars. In this paper we evaluate the detection significance of compensated isocurvature perturbations (CIPs) from observations of the 21-cm hydrogen-line during the cosmic-dawn era. CIPs are modulations of the relative baryon and dark-matter density that leave the total matter density unchanged. We find that, under different assumptions for feedback and foregrounds, the ongoing HERA and upcoming SKA1-low experiments will provide constraints on uncorrelated CIPs at the level of $\sigma(A_{\rm CIP})= 10^{-3}-10^{-4}$, comparable to the sensitivity of upcoming CMB experiments, and potentially exceeding the constraints from cosmic-variance limited BAO surveys. 

\end{abstract}

\maketitle

\section{Introduction}

% \footnotetext[$^a$]{: Equal contributions.}
%\makeatletter{\renewcommand*{\@makefnmark}{}
% \footnotetext{\hspace{-0.2cm}$^a$ equal contributions.}\makeatother}

The standard cosmological paradigm ($\Lambda$CDM) with single-field inflation~\citep{Starobinsky:1979ty,Starobinsky:1980te,Kazanas:1980tx,Sato:1980yn,Guth:1980zm} predicts adiabatic initial conditions with inhomogeneities in DM, baryons, neutrinos, and photons all uniquely determined by the primordial curvature perturbations~\citep{Weinberg:2008zzc,Weinberg:2003sw,Weinberg:2004kr,Weinberg:2008nf}. More general theories with multiple degrees of freedom, however, can source non-adiabatic (isocurvature) perturbations, where the relative mixture of DM, baryons, neutrinos, and photons become independent degrees of freedom~(see e.g.~\citep{Weinberg:2004kf,Weinberg:2008si}). While measurements of the CMB and galaxy distributions put tight constraints on most forms of isocurvature perturbations, a specific form of isocurvature perturbation is difficult to constrain from CMB and galaxy surveys alone: compensated isocurvature perturbations (CIPs). CIPs are fluctuations of baryons and DM that leave the total matter perturbations unchanged and adiabatic. Since the CMB is only sensitive to the total matter fluctuations, at leading order in the perturbation amplitude, CIPs avoid stringent constraints from measurements of the CMB alone, allowing for CIPs to have an amplitude orders of magnitude larger than the adiabatic modes~\citep{Holder:2009gd,Gordon:2009wx,Grin:2011tf,Grin:2011nk,Smith:2017ndr, Munoz:2015fdv,Grin:2013uya,Heinrich:2019sxl,Hotinli:2019wdp}. A detection of CIPs can provide insight into both the number of primordial fields that contribute to the observed density fluctuations, as well as their decay channels (see, e.g., Refs.~\citep{He:2015msa,Smith:2015bln}).  

Regardless of whether adiabatic or isocurvature, primordial fluctuations seed the rich large-scale structure of matter we observe in our Universe. As matter clusters under gravity, however, its components can behave very differently. While the majority of matter is collisionless, dark, and cold; a fraction of it are baryons which couple to photons before recombination (at redshift $z\simeq1100$), giving rise to the baryon acoustic oscillations (BAOs) observed in the CMB and galaxy surveys. The same physics also induces a bulk relative velocity between DM and baryons~\citep{Tseliakhovich:2010bj},
which strongly affects the formation of the first stars during the cosmic dawn era ($z\simeq10-30$)~\citep{Dalal:2010yt,Naoz:2011if,Tseliakhovich:2010yw,Greif:2011iv,McQuinn:2012rt,Stacy:2010gg,Naoz:2011if,Fialkov:2011iw,Yoo:2011tq,Pritchard:2011xb,Barkana:2016nyr,Munoz:2019rhi} that will be soon accessed by measurements of the 21-cm hydrogen line. During these early times the typical mass of collapsed baryonic objects fall near the critical mass below which gas pressure prevents their collapse. The abundance of Lyman-$\alpha$ photons that excite the hyperfine transition in neutral hydrogen, and allow it to absorb 21-cm photons from the cosmic microwave background (CMB), will depend on the collapsed fraction of baryons.
Thus, it is directly impacted by effects that alter early structure growth such as local modulations of the relative velocity between dark matter (DM) and baryons, which imparts the acoustic scale onto the signal.

It was shown in Ref.~\citep{Munoz:2019fkt} that the acoustic signature from the relative velocities takes the form of velocity acoustic oscillations (VAOs), whose shape is unaffected by astrophysics and can be used as a standard ruler.
These VAOs can be observed with upcoming 21-cm power-spectrum experiments such as HERA~\citep{DeBoer:2016tnn} or SKA1-low~\cite{2019arXiv191212699B,Bacon:2018dui}. The VAO feature provides an effective probe of the early-Universe physics that affect the relative behaviour of DM and baryons. 
In this study, we will use the VAOs to look for CIPs.

Unlike the usual BAOs in the matter power-spectrum (whose amplitude is small) the VAO feature is $\mathcal{O}(1)$ in the 21-cm power spectrum~\citep{Munoz:2019rhi}. Furthermore, as is the case for the BAO feature in the CMB and LSS observables, some characteristics of the VAOs are unaffected from the complicated local physics related to various feedback mechanisms which play a role during the epoch of reionization, and can be utilised to constrain effects that have a coherent impact on the observables on large-scales, such as CIPs. In this paper we discuss the detection significance of CIPs from measurement of the 21-cm hydrogen line.

This paper is organised as follows. In Section~\ref{sec:the_cips} we discuss CIPs and their effect on the relevant observables, and discuss the sensitivity of the 21-cm power spectra on the effect of CIPs. In Section~\ref{sec:21cm_ps}  we review the 21-cm hydrogen line and the effect of VAOs on its power spectrum. In Section~\ref{sec:CIPreconstruction}, we evaluate the detection significance of CIPs using both the full shape of the power spectra as well as change in the VAO scale. For the latter we introduce as a robust measure by marginalising over the parameters that describe the smooth part of the 21-cm power spectra and the VAO feature. We conclude with discussion in Section~\ref{sec:discussion}. We describe our noise calculations in Appendix~\ref{app:telescopes}.

\section{CIPs and their effects on observables}\label{sec:the_cips}

Theoretical models of the early Universe such as inflation with multiple degrees of freedom can naturally give rise to isocurvature perturbations.  Isocurvature perturbations can be parameterised by the fractional number-density difference between photons and other species as, 
\be
S_{i\gamma} \equiv \frac{\delta {n}_i}{\bar{n}_i}-\frac{\delta n_\gamma}{\bar{n}_\gamma}\,,
\ee
where $\gamma$ is for photons, $\bar{n}_i$ is the unperturbed number density and $\delta n_i$ is the number-density fluctuation, with $i=\{b,c,\nu\}$ for baryons, DM, and neutrinos, respectively. We define CIPs with an amplitude $\Delta$ as having related baryon- and CDM-isocurvature perturbations with 
\be\label{eq:iso_cips_definition}
S_{b\gamma}=\Delta\,\,\ \ {\rm and} \ \ \,\,S_{c\gamma}=-\frac{\rho_b}{\rho_c}\Delta\,,
\ee
where $\rho_i$ is the energy density of species $i$. 

CIPs can be parameterised with a scale-invariant power spectrum, for example, as studied in Ref.~\citep{Smith:2017ndr}. Depending on the sourcing physical process, CIP fluctuations can either be correlated or uncorrelated with the adiabatic curvature fluctuations $\zeta$. In the former case, cross-correlating the reconstructed CIP field with the underlying density fluctuations significantly improves the detection prospects of CIPs and allow using sample-variance cancellation techniques upon cross-correlating different tracers (such as the bulk velocity fluctuations reconstructed from the measurements of the kSZ effect, as studied in \citep{Hotinli:2019wdp}).\footnote{CIPs may be sourced, for example, in the curvaton inflation scenario~\citep{Lyth:2001nq,Lyth:2003ip} - a spectator scalar field which is subdominant in the early Universe (with respect to the inflaton field that is driving the inflationary dynamics) and significantly contributes to the curvature fluctuations after the end of inflation.} 
In the case of \textit{uncorrelated} CIPs which we consider in this paper, cross-correlations cannot be utilised to improve the signal-to-noise, resulting in significantly more pessimistic detection prospects. It is hence important to find independent ways of measuring uncorrelated CIPs. 

More generally, since the primordial CIPs could be sourced by the gravitational potential in the early Universe, they can constitute to a significant source of density differences between baryons and DM on large scales. These observational signatures of primordial CIPs are largely protected from complicated non-linear physics due to the equivalence principle, which dictate that it is difficult for local interactions to produce coherent effects on large scales.

For CIPs with small amplitudes and long wavelengths that exceed the sound horizon, observable implications of varying fractional baryon and DM number density can be captured by the separate-universe approximation around a patch at some location $\boldsymbol{X}$, with perturbed cosmological parameters,
\be
\delta {\Omega}_b=\bar{\Omega}_b\,\Delta(\boldsymbol{X})\,\ \ {\rm and}\ \ \,\delta {\Omega}_c=-\bar{\Omega}_b\,\Delta(\boldsymbol{X})\,,
\ee
where $\delta\Omega_b$ ($\delta\Omega_c$) is the CIP-induced modulation to the baryon (DM) fluctuations and the overbar represents their unperturbed $\Lambda$CDM values. As a consequence, the sound speed of the baryon-photon fluid around $\boldsymbol{X}$ changes as
\be
c_s\rightarrow (1+[3\bar{\rho}_b(1+\Delta(\boldsymbol{X}))]/4\bar{\rho}_\gamma)^{-1/2}\simeq{\alpha}(\boldsymbol{X})c_s\,,\non
\ee
where $\alpha(\boldsymbol{X})=1+\Delta(\boldsymbol{X})/C$ and $C\equiv-2(1+R)/R$ with $R\equiv3\bar{\rho}_b/4\bar{\rho}_\gamma$ and $C\simeq-5.23$ for standard cosmology~\citep{Heinrich:2019sxl}.
The change in the BAO scale due to CIPs is then 
\be 
r_{\rm drag}(\boldsymbol{X})\rightarrow\alpha(\boldsymbol{X})\,r_{\rm drag}\,,
\label{eq:rsshift}
\ee
leading to the modulation of the relative-velocity power spectrum (which we introduce next) in the form $\Delta^2_{21,\rm vel}(k,z;\boldsymbol{X})\rightarrow\Delta^2_{21,\rm vel}(\alpha(\boldsymbol{X})k,z)$. Note, however, that the dominant effect of CIPs on the 21-cm power spectrum is modulating its amplitude by locally altering the baryon density, on which the brightness temperature depends directly via the baryon fraction.

We show the effect of CIPs on the 21-cm power spectrum in Figure~\ref{fig:display_CIPs}. Differently from CMB and LSS observables, the direct dependence of the 21-cm amplitude to baryon fraction also provides a unique and potentially powerful probe of CIPs.

\section{The impact of relative velocities on the 21-cm power spectrum}\label{sec:21cm_ps}

\subsection{The 21-cm hydrogen line}

The hyperfine splitting of the ground state of neutral hydrogen possesses a triplet and singlet ground state. These two states are spin-flip states and the forward (backward) transition from the triplet state to singlet state is accompanied by an emission (absorption) of a 21-cm wavelength photon. 
Whether cosmological hydrogen emits or absorbs 21-cm photons can be understood by calculating its spin temperature. 
The spin temperature $T_s$ local to the hydrogen can be described by the relation 
\be
\frac{n_1}{n_0}=\frac{g_1}{g_0}e^{-T_*/T_s}\,,
\ee
where $n_0$ ($n_1$) is the comoving number density of the hydrogen atoms in the singlet (triplet) state, $g_0=1$ ($g_1=3$) are their numbers of degrees of freedom, and $T_*=0.068$\,K is the temperature corresponding to the 21-cm hyperfine transition. 
Observations are made in reference to the CMB.
When the local spin temperature is higher (lower) than the CMB temperature, hydrogen emits (absorbs) photons from the CMB. 
The distribution of these photons at the different wavelengths can be studied to understand the astrophysics and cosmology of our Universe at different redshifts. 
The main observable of interest is the 21-cm brightness temperature~\citep{2006PhR...433..181F},
\be\label{eq:21cm_temperature}
T_{21} = 38{\rm mK}\left(1-\frac{T_\gamma}{T_s}\right)\left(\frac{1+z}{20}\right)^{\!1/2}\!\!x_{\rm HI}(1+\delta_b)\frac{\partial_rv_r}{H(z)}\,,\non
\ee
where $x_{\rm HI}$ is the neutral-hydrogen fraction and $\partial_r v_r$ is the line-of-sight gradient of the velocity, $T_\gamma$ is the CMB temperature, $\delta_b$ is the baryon overdensity and $H$ is the Hubble parameter. For a review of the 21-cm line see e.g \citep{Barkana:2016nyr,Pritchard:2011xb,2006PhR...433..181F}.

The epoch of interest to the study in the paper is the cosmic-dawn era, defined by the formation of the first stars, theorised to begin around $z \sim 25-35$~\citep{2014MNRAS.437L..36F}. %($z\!\gtrsim\!15$), 
Initially, after recombination, the gas kinetic temperature is dominated by its adiabatic cooling.
The high density couples the spin temperature to that of the CMB background through collisions~\citep{Loeb:2003ya}.
As the Hubble flow progresses, collisional coupling of hydrogen becomes ineffective and the 21-cm signal vanishes at the end of the dark ages.
During the cosmic dawn the first stars will produce a UV background, which will redshift into the Lyman-$\alpha$ line and couple the spin and kinetic temperatures of hydrogen in the intergalactic medium (IGM) via the Wouthuysen-Field effect \citep{1952AJ.....57R..31W, 1958PIRE...46..240F,Hirata:2005mz}.
Remnants of these first stars are likely to produce a diffuse background of $\sim 0.1 - 2 ~\rm keV$ X-rays~\citep{Pacucci:2014wwa,Pritchard:2006sq}, 
heating the IGM before reionization progresses largely after $z \sim 10$~\citep{Morales:2009gs,Mesinger:2018ndr,Wise:2019qtq}. As the tail-end of reionization is approached ($z<10$), the effects of streaming velocities in the IGM on the 21-cm signal are reduced by Lyman-Werner feedback \citep{2014MNRAS.437L..36F,FGU}.

CIPs can affect the 21-cm signal in a multitude of ways, some of which are degenerate with as of yet unknown astrophysics \citep{2009PhRvD..80f3535G}, and therefore observations benefit from utilizing the VAO signature. For our observation we adopt a redshift range of $10<z<25$ so that the bulk of the cosmic dawn is occurring but the IGM streaming velocities are yet to become small.

We calculate the observable signal using the semi-numerical simulations provided by \texttt{21cmvFAST}\footnote{\href{https://github.com/JulianBMunoz/21cmvFAST}{github.com/JulianBMunoz/21cmvFAST}}, which is built upon  \texttt{21cmFAST}\footnote{\href{https://github.com/andreimesinger/21cmFAST}{github.com/andreimesinger/21cmFAST}}. 
Initial conditions for density and peculiar velocity fields are set at $z=300$ with a Gaussian random field in Lagrangian space, before being evolved with the Zel'dovich approximation \citep{1970A&A.....5...84Z} to match the mean collapse fraction for the conditional Sheth-Tormen halo mass function \citep{1999MNRAS.308..119S}.
The sources embedded in each halo are assumed to emit photons at a rate proportional to the increase of the total collapsed halo mass (for a different parametrization see Refs.~\cite{Park:2018ljd,Qin:2020xyh}). 
In each cell, the excursion set formalism is used to estimate the mean number of sources contributing to the gas temperature from the surroundings.
The kinetic temperature is calculated including adiabatic expansion, Compton scattering with the CMB~\citep{2011ascl.soft06026S}, and the inhomogeneous heating history of the IGM (through a combination of X-rays and collisional coupling). Details on this calculation can be found in Refs.~\citep{2011MNRAS.411..955M,Murray:2020trn}.

We produce realizations of the 21-cm signal in 2000 Mpc boxes on 2000$^3$ grids of coeval cubes for each observed frequency. We simulate one coeval cube at the respective redshift from an initial density field given by appropriate transfer functions for matter and relative velocities. 
The quantity we are interested in is the power spectrum of the 21-cm signal which can be written as 
\be
\langle \delta T_{21}(\vec{k},z) \delta T_{21}^*(\vec{k}',z) \rangle = (2\pi)^3\delta(\vec{k}-\vec{k}')P_{21}(\vec{k},z)\,,
\ee
where $\delta T_{21}(\vec{k},z)$ is the Fourier transform of $[ T_{21}(\vec{x}) - \bar{T}_{21} ] / \bar{T}_{21}$, the zero-mean fluctuations of the 21-cm brightness temperature at redshift $z$. 
We define the spherically averaged power spectrum as 
\be
\Delta_{21}^2=P_{21}(k,z)\dfrac{k^3}{2\pi^2},
\ee
which we will often refer to as the 21-cm power spectrum for convenience.

\begin{figure}[b]
    \vspace{-0.25cm}
    \includegraphics[width=\columnwidth]{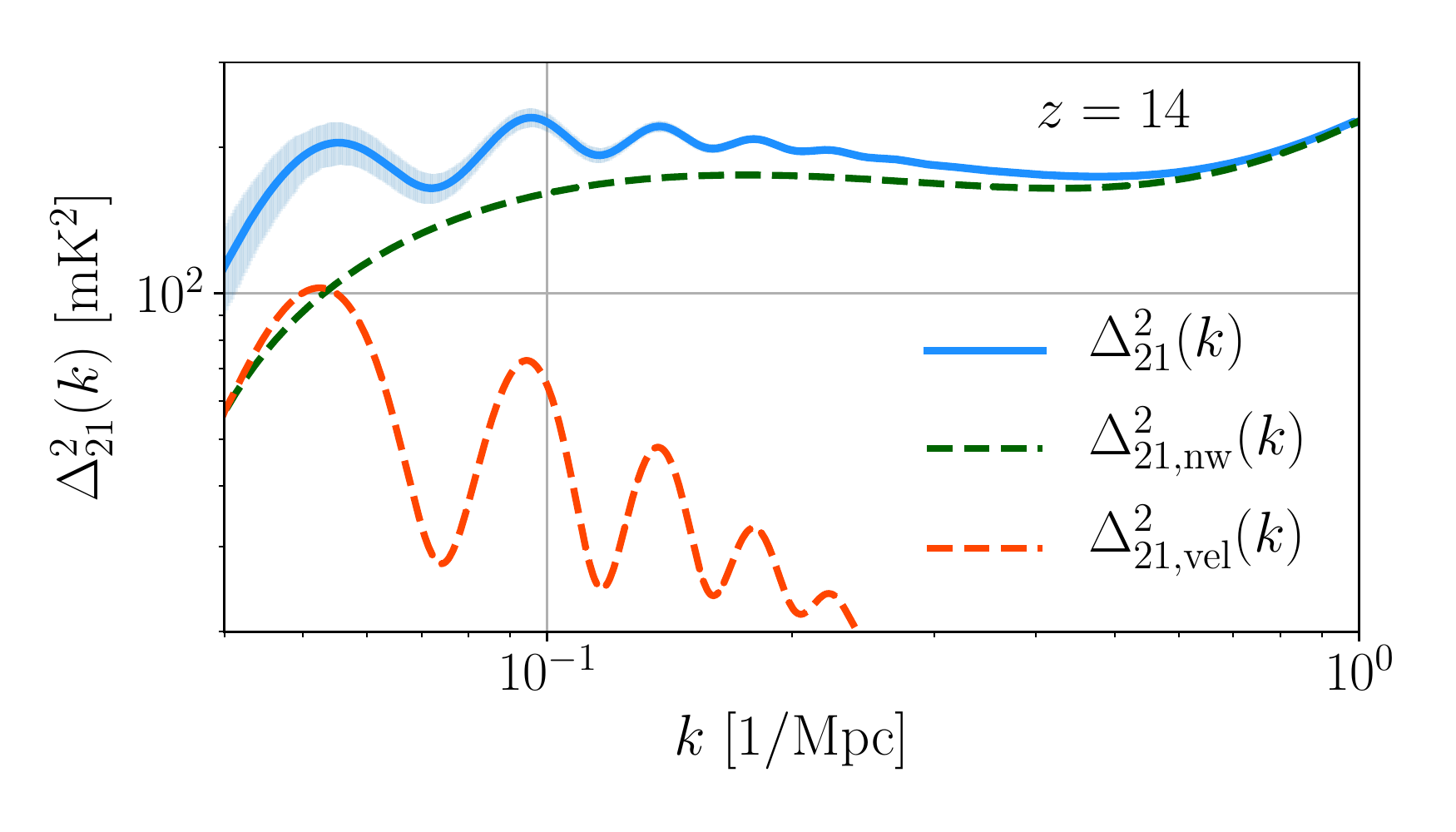}
    \vspace{-0.75cm}
    \caption{The 21-cm hydrogen line brightness-temperature power spectra at $z=14$ (in solid blue), shown with fitted smooth contribution to the power spectra (in dashed green) in the absence of relative velocity fluctuations and the velocity contribution (dashed orange) as discussed in Section~\ref{sec:impact_of_v}. We have considered medium feedback and used \texttt{21cmvFAST}~\citep{Munoz:2019fkt}. Error bars shown in the figure for the signal are Poisson errors from our simulations. Dashed green lines are fourth-order polynomial fits to the brightness temperature spectra, as discussed in Section~\ref{sec:impact_of_v}.}
    \label{fig:display_CIPs_VAOs}
\end{figure}

\subsection{The impact of relative velocities}\label{sec:impact_of_v}

The modulation of the 21-cm power spectrum due to DM-baryon relative velocities can be captured to a good approximation from the statistics of the collapsed baryonic density. 
In short, the effect of bulk relative velocities is akin to that of the gas pressure, suppressing the accretion of baryons. As the gas falls into the DM halo, its bulk kinetic energy is converted into thermal energy resulting in a change in the effective sound speed $c_{\rm eff,s}\simeq(c_s^2+v_{\rm cb}^2)^{1/2}$, hence in the critical mass scale and in the baryon collapsed fraction~\citep{Dalal:2010yt,Naoz:2011if,Tseliakhovich:2010yw,Greif:2011iv,McQuinn:2012rt,Stacy:2010gg,Naoz:2011if,Fialkov:2011iw,Yoo:2011tq,Pritchard:2011xb,Barkana:2016nyr,Munoz:2019rhi}.
The effect of the relative velocities on the amplitude of the 21-cm brightness temperature power spectrum can then be parameterised as~\citep{Munoz:2019fkt}
\be\label{eq:velo_pert}
\Delta_{21,\rm vel}^2(k,z)=A_{\rm vel}(z)\Delta_{v^2}^2(k,z)|W(k,z)|^2\,,
\ee
where $A_{\rm vel}$ is some redshift-dependent amplitude of fluctuations.
The window function, $W(k,z)$, can be utilised to isolate the different contributors to the 21-cm power spectrum such as Lyman-$\alpha$ coupling and X-ray heating. 
We defined $\Delta_{v^2}^2(k)$ as the power spectrum of the quantity 
\be
\delta_{v^2}=\sqrt{\frac{3}{2}}\left(\frac{v_{\rm cb}^2}{v_{\rm rms}^2}-1\right)\,,
\ee
which accurately captures the shape of the effect of relative velocities on the observables for the scales of interest where the `streaming' bulk relative velocity can be approximated with a root-mean-squared value $v_{\rm rms}\simeq30 \,{\rm km\,s}^{-1}$ at recombination. Note that the coefficient $A_{\rm vel}$ is a model-dependent amplitude that is not directly observable, similar to the BAO amplitude.  As the VAOs are statistically independent from the density fluctuations at first order, the amplitude of the 21-cm power spectrum can be written as
\be\label{eq:vao_definition}
\Delta_{21}^2(k,z)=\Delta_{21,\rm vel}^2(k,z)+\Delta^2_{21,\rm nw}(k,z)\,,
\ee
where $\Delta^2_{21\!,\rm nw}(k,z)$ is the component \textit{excluding} VAOs. Throughout this paper we parameterise the smooth contribution to the spectra as a fourth-order polynomial following Ref.~\cite{Munoz:2019rhi},
\be\label{eq:smooth_pw}
\ln[\Delta_{21,\rm nw}^2(k,z)]=\sum_{i=0}^{4}c_i(z)[\ln k]^i\,,
\ee
where $c_i(z)$ are coefficients we fit for, using simulations we discuss in Section~\ref{sec:sims}. The fitted smooth spectra serve as our phenomenological model whose parameters we marginalise in our forecasts. We model the velocity power spectrum as in Ref.~\citep{Munoz:2019fkt} using the form we defined in Eq.~\eqref{eq:velo_pert}. We calculate the window function and the amplitude $A_{\rm vel}(z)$ for a given feedback model using  \texttt{21cmvFAST}, and calculate $\Delta_{v^2}^2$ for a given cosmology. Later in our forecasts we will take $A_{\rm vel}(z)$ as a free parameter to capture the model dependence of the VAO amplitude to the complicated baryonic physics. We display the effect of VAOs on the 21-cm hydrogen line in Figure~\ref{fig:display_CIPs_VAOs}.

\section{CIP reconstruction}\label{sec:CIPreconstruction}

\subsection{Simulations}\label{sec:sims}

We use the \texttt{21cmvFAST} software to simulate the brightness temperature from co-eval boxes of size 2000 Mpc on 2000$^3$ grids in the redshift range, $z\in[4,30]$. We produce 20$\times3$ simulations with different initial conditions and for three considerations of baryonic-feedback levels, defined as low, medium and high in \texttt{21cmvFAST} settings, to observe how the astrophysics of cosmic dawn alters the effect of VAOs on the 21-cm brightness temperature power spectra. All simulations have the same cosmological volume, grid space and redshift range. For each baryonic feedback level, we produce the brightness temperature power spectra from each co-eval box and average over the 20 realizations with different initial conditions to acquire a theory prediction for the spectra. The spectra from simulations is subject to (Poisson) simulation shot-noise and sample variance. We use these averaged spectra as the signal, throughout. Using the separate-universe approximation, we repeat the calculation for three different levels of CIPs; taking $\Delta\in\{-0.05,0.0,0.05\}$. This amounts to producing simulations with different baryon and DM densities, appropriate to the CIP amplitude, and the same total matter density. In total, we produce $3\times3$ power spectra for each coeval box at a given redshift, for three CIP amplitudes and three feedback levels. For a given feedback level and redshift, the three different power spectra represents separate universes with different levels of CIPs.\footnote{Note that we omit the effect of CIPs on the collapsed fraction in  our simulations (i.e. the \texttt{Fcoll} tables in \texttt{21cmvFAST}), which may futher enhance the sensitivity of the 21-cm hydrogen line to the CIPs. 
}

\subsection{Measurement errors}

We calculate the anticipated experimental noise for HERA and SKA1-low using the software~\texttt{21cmSense}\footnote{\href{https://github.com/jpober/21cmSense}{github.com/jpober/21cmSense}}~\citep{Pober:2012zz, Pober:2013jna,  2016ascl.soft09013P}, which we describe in~Appendix~\ref{sec:noise_calc}. In order to forecast noise (and the cosmic variance) in sub-volumes (boxes) centred at $\boldsymbol{X}$, we modify the mode-integral in this code to introduce a cut-off on Fourier modes larger than the size of our boxes, i.e., we force $k_\parallel,k_\perp>2\pi/r_{\rm box}$, and rescale the volume seen by each mode at a given redshift (and a given bandwidth) to the appropriate volume of a given box. We choose varying box sizes in the range $r_{\rm box}\in[150$Mpc,1500Mpc$]$ all of which are significantly smaller than the total survey sizes of the HERA and SKA1-low experiments we consider. We take these boxes as a proxy for separate universes with varying CIP amplitude. Note that for most of the box sizes, the SNR for detecting fluctuations is much lower than that is for the total volume. Next, we measure the CIP amplitude in each subvolume and reconstruct the survey-wide large-scale CIPs by combining all boxes that fit into our survey volume for a given redshift range.

\subsection{Reconstruction}\label{sec:reconstruction}

Here we describe our VAO reconstruction procedure, similar to that of Ref.~\citep{Heinrich:2019sxl}. We evaluate the detection significance of CIPs for a fixed cosmology, assuming that a 21-cm hydrogen-line survey can locally test the observed spectra against the effect of CIPs inside different boxes centred at $\boldsymbol{X}$, and of size smaller than our simulation box and the survey (and larger than the mean free path of X-ray and UV photons during cosmic dawn). 
The local CIP measurements from the 21-cm data can be biased due to our poor understanding of the underlying astrophysics.
For example, feedback processes can change the amplitude of the signal (parametrised through $c_i$ in our Eq.~\eqref{eq:smooth_pw}) and will be degenerate with CIPs, contributing a large theoretical uncertainty.
Such degeneracies can potentially be surmounted by external measurements of the same astrophysics (e.g.~through galaxy UV luminosity functions~\citep{Tacchella:2018qny,Gillet:2019fjd,Sabti:2020ser}) or by careful modelling, in which case the model parameters need to be marginalised, weakening the constraining power of the 21-cm data.

We model the smooth part of the 21-cm brightness temperature power spectra as given in Eq.~\eqref{eq:smooth_pw}, taking the five coefficients of the polynomial as free parameters at each redshift. We calculate the fiducial values for $W(k,z)$,  $\Delta_{v^2}$ and $A_{\rm vel}(z)$ [as defined in Eq.~\eqref{eq:velo_pert}]. We take the former (VAO amplitude) as model parameter and set the first two quantities fixed. Together with the CIP amplitude, our phenomenological model involves 7 parameters: 
\be\label{eq:parameters}
\{c_0(z),c_1(z),c_2(z),c_3(z),c_4(z),A_{\rm vec}(z),\Delta\}\,,
\ee
for each box centred at some redshift $z$.

For each given box, we define the Fisher information matrix as 
\begin{equation}
\label{eq:fisherM}
F_{\alpha\beta}(z) = \sum\limits_{k-\rm bins}\frac{\partial_\alpha\Delta_{21}^2(k,z)\partial_\beta\Delta_{21}^2(k,z)}{(\Delta_{21,\rm obs}^{2})^2(k,z)}\,,
\end{equation}
where $\{\alpha,\,\beta\}$ vary over the parameters in Eq.~\eqref{eq:parameters}, $\sum_{k\!-\!\rm bin}$ is the sum over the binned wavenumbers and $\Delta_{21,\rm obs}^{2}(k)$ is the total variance (including the thermal and the cosmic-variance noise) for a given $k$-bin, which we calculate with \texttt{21cmSense} (see more details in~Appendix~\ref{sec:noise_calc}). The error on the local CIP amplitude for an individual box can then simply be written as
\be
\sigma_{\Delta}(z)=\sqrt{[F^{-1}(z)]_{\Delta\Delta}}\,.
\ee

Next, we use the separate-universe approximation in each box to reconstruct the large scale CIP fluctuations in Fourier space. The latter can be estimated from the Fourier transform of the locally measured CIP amplitude $\Delta(\boldsymbol{X})$. We assume sufficiently many boxes can be utilised for constraining CIPs; hence the effect of dividing the survey volume to smaller parts can be approximated by writing the reconstructed CIP field as a convolution of the true field, $\Delta(\boldsymbol{k},z)$, with a radial tophat window function in real space, which takes the form, 
\be
W(kr)\equiv 3[\sin(kr)-kr\cos(kr)]/(kr)^3\,,
\ee
in Fourier space. The sensitivity on the local CIP amplitudes can be related to to those of CIP fluctuations in Fourier space in the full survey volume as
\be
\label{eq:recon_noise}
N_{\Delta\Delta}^{\rm rec}(k,z)\equiv \Lambda(z)\,[W(k r)]^{-2}\,,
\ee
where $\Lambda(z)\equiv\sigma_{\Delta}^{2}(z)V_{\rm box}$ is approximately independent of the box volume for boxes whose size is sufficiently smaller than the total survey and $r=r_{\rm box}$ is the size of the boxes where the local CIP amplitudes are measured. Finally, we define the scale-invariant CIP power spectra as 
\be\label{eq:CIPsignal}
P_{\Delta\Delta}(k)\equiv A_{\rm CIP}k^{-3}\,.
\ee 
The error on $A_{\rm CIP}$ can then be calculated as 
\be\label{eq:recon_noise2}
\sigma_{A_{\rm CIP}}^{-2}=\frac{1}{A_{\rm CIP}^2}{V_{\rm bin}}\int\frac{\dd k\,k^2}{2\pi^2}\left(\frac{P_{\Delta\Delta}(k)}{N_{\Delta\Delta}^{\rm rec}(k,z)}\right)^2\,,
\ee 
where $V_{\rm bin}$ is the total survey volume inside the redshift bin and the integral over the Fourier wavenumber is bounded by the size of the box (redshift bin) on small (large) scales.

\subsection{Other effects and priors}\label{sec:priors}

\begin{figure}[t]
    % \hspace*{-0.8cm}
    \includegraphics[width =1\columnwidth]{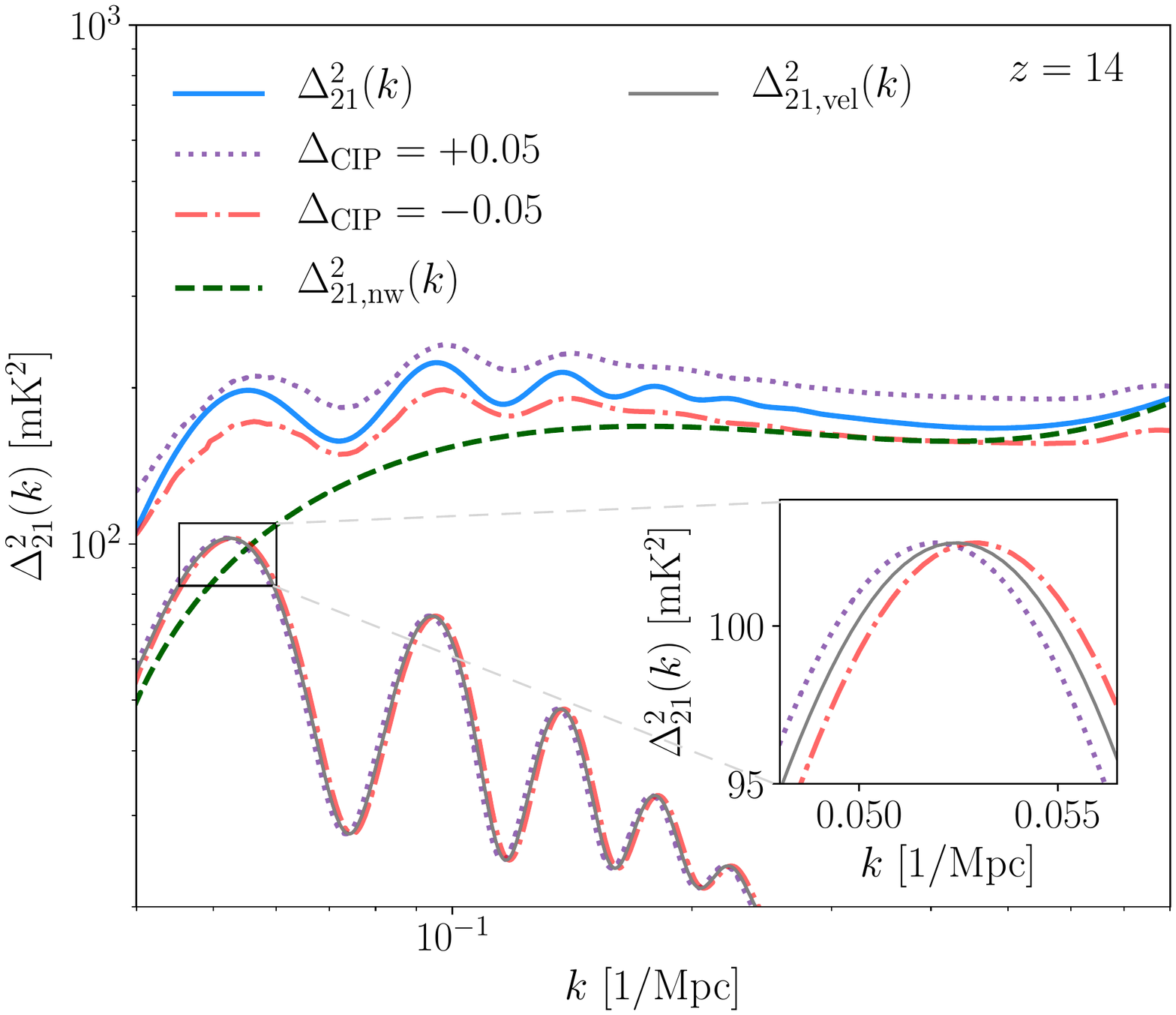}
    \vspace{-0.5cm}
    \caption{The effect of locally varying the baryon-DM ratio and the speed of sound $c_{s}$ of the baryon-photon plasma due to CIPs on the VAO signature and brightness temperature power spectrum, at redshift $z\!=\!14$. The effect of velocities were calculated with the medium-feedback assumption using \texttt{21cmvFAST} software. The blue solid line shows the 21-cm power spectrum (fitted from simulations) with zero CIP amplitude, $\Delta\!=\!0$. The red and violet dotted/dashed lines shows the power spectrum in a universe where $\Delta\!=\!\pm0.05$. Similar to Figure~\ref{fig:display_CIPs_VAOs}, the green dashed line is the fitted temperature power spectrum excluding the effect of VAOs. The gray solid line shows the contribution of the relative-velocity effect discussed in this paper, in the absence of CIPs. The orange (purple) dot-dashed (dotted) lines show the effect of CIPs on the relative velocity power spectrum, which shift the VAO peaks by locally modulating the acoustic scale $r_{\rm drag}$ by Eq.~\eqref{eq:rsshift}. Both this shift and the overall amplitude change are included in our forecasts.
    }
    \label{fig:display_CIPs}
\end{figure}

Parameters that define the smooth spectra and the VAO amplitude depend on cosmology and the astrophysics of the cosmic-dawn era. Much about their dependence on the latter is yet unknown. Furthermore, processes such as local baryonic feedback could vary spatially depending on the characteristics of the involved mechanisms and the influence of bulk fluctuations, which can potentially be confused with the effect of CIPs (as shown in Figure~\ref{fig:display_CIPs}). In order to isolate the effect of CIPs on the 21-cm power spectra, we must then marginalise over the non-CIP parameters locally (inside each box). 

We assume that, in the limit of large boxes and random, uncorrelated, distribution of parameters ($c_i(z)$ and $A_{\rm vec}(z)$), the error on the parameters are dominated by the measurements (including both cosmic-variance and thermal noise). If the mapping between the parameters and the power spectra is unchanged between boxes (in the absence of CIPs), then those can be considered nuisance parameters which can be measured from the full survey volume. 
In such a case, one could use the Fisher matrix from the full survey volume as a prior on the non-CIP parameters, in the form $F^{\rm box}_{\alpha\beta}(z)\rightarrow F^{\rm box}_{\alpha\beta}(z)+F^{\rm global}_{\alpha\beta}(z)$, where $\alpha,\,\beta$ vary over the non-CIP parameters. 

In practice, however, the spatial variations of this mapping may depend on many factors, which need to be modelled and studied with simulations. As it is technically challenging to predict how the 21-cm spectra vary locally over the survey volume, we instead vary the 21-cm power-spectrum amplitude locally, with fluctuations in a range of $1\%$ to $10\%$ of the globally measured spectra, $\bar{P}_{21\rm cm}$.\footnote{The local variation of the 21-cm power spectrum has been discussed in the recent literature (see e.g.~\citep{Giri:2018dln} where authors find a percent-level fractional change in the local 21-cm power spectra on scales $k\in[0.1,3]$/Mpc, inside boxes of size $\sim\!700$Mpc$^3$ and redshifts $z\in[6,12]$). Note that the influence of bulk effects can be collectively parametrised with a squeezed bispectra, as discussed in \citep{Giri:2018dln}, and can potentially be measured to improve the detection significance of CIPs and other effects.} For each choice of variation, we have generated $10^4$ random realizations of the power spectra and fitted the 6 non-CIP parameters to calculate their covariance matrix, $\mathcal{C}_{\alpha\beta}^{\rm local}$, in the presence of local spatial variations. We then use this as a limit on how well the non-CIP part of the power spectra can be measured at each box, by transforming the non-CIP part of the Fisher matrix in the form
\begin{equation}
    F_{\alpha\beta}^{\rm box}\rightarrow {F}_{\alpha\beta}^{\rm box}+[(F^{\rm global}_{\alpha\beta})^{-1}+\mathcal{C}_{\alpha\beta}^{\rm local}]^{-1}, 
\end{equation}
where the Fisher matrix with global superscript on the right-hand side includes the information from the full survey volume as described above. This effectively includes a global prior on the non-CIP parameters ($F^{\rm global}_{\alpha\beta}$), subject to the constraint that these vary from box to box (following $\mathcal{C}^{\rm local}_{\alpha\beta}$).

Our results are sensitive to this intrinsic fluctuation amplitude (as shown in Figure~\ref{fig:display_CIPs_hera_ska}  for the upcoming HERA and SKA1-low experiments and for moderate and optimistic foreground considerations). Next, we discuss the details of our forecast.

\subsection{Forecasts}\label{sec:forcasts}

We take four redshift bins of size $\Delta z=2$, centred at redshifts $z\in\{12,14,16,18\}$, and one redshift of size $\Delta z=3$, centred at $z=24$. We describe the experimental survey specifications we consider in Appendix~\ref{app:telescopes}. For reference, we find the total (over all $z$) detection signal-to-noise ratio (SNR) of the 21-cm signal to be $\{130,190,498\}$, and the SNR of the VAO signature to be $\{21,40,190\}$, for our $\{$pessimistic,\,moderate,\,optimistic$\}$ foreground considerations, respectively, and for regular baryonic feedback, using the specifications for the HERA survey.\footnote{We calculate the total SNR as equal to $\sum_k\!\Delta^2_{21}(k,z)/\Delta_{21}^{2,\rm obs}(k,z)$ and the VAO SNR as equal to $\sum_k\!\Delta^2_{21,\rm vel}(k,z)/\Delta_{21}^{2,\rm obs}(k,z)$, summed (in  quadrature) over the redshift bins we consider. $\sum_k$ is the sum over binned wavenumbers as described in the text.} For the same feedback and foreground choices, we find SKA1-low SNR equal to $\{$80,\,133,\,800$\}$ for the total signal, and $\{$14,\,29,\,311$\}$ for the VAOs. 

We define three levels of foreground contamination: optimistic, moderate and pessimistic, as described in Appendix~\ref{sec:noise_calc}. We show forecasts in Figure~\ref{fig:display_CIPs_hera_ska} for HERA and SKA1-low surveys. The upper lines in each plot correspond to pessimistic assumptions for foreground contamination, while the middle blue and lower orange lines correspond to moderate and optimistic foreground contamination assumption, respectively. Figure~\ref{fig:display_CIPs_hera_ska} suggest that even the first generation 21-cm surveys, such as HERA and SKA1-low, can provide constraining power comparable to stage 3 and 4 large-scale structure surveys and potentially reach levels comparable to cosmic variance limited (CVL) CMB experiments, depending on the foreground and the modelling uncertainties of the VAOs and the smooth component of the power spectra. 
For our standard case of medium feedback and moderate foregrounds, we find that CIPs can be detected at the $95\%$ confidence level (CL) in HERA if ${A_{\rm CIP}} > 2.3\times10^{-3}$ and in SKA1-low if ${A_{\rm CIP}}> 6.3\times10^{-3}$, sensitivities comparable to those from large-scale structure surveys alone.

\subsection{Summary of this section}

\begin{figure*}[ht!]
\centering
\begin{subfigure}{}
  \centering
  \includegraphics[width=.46\linewidth]{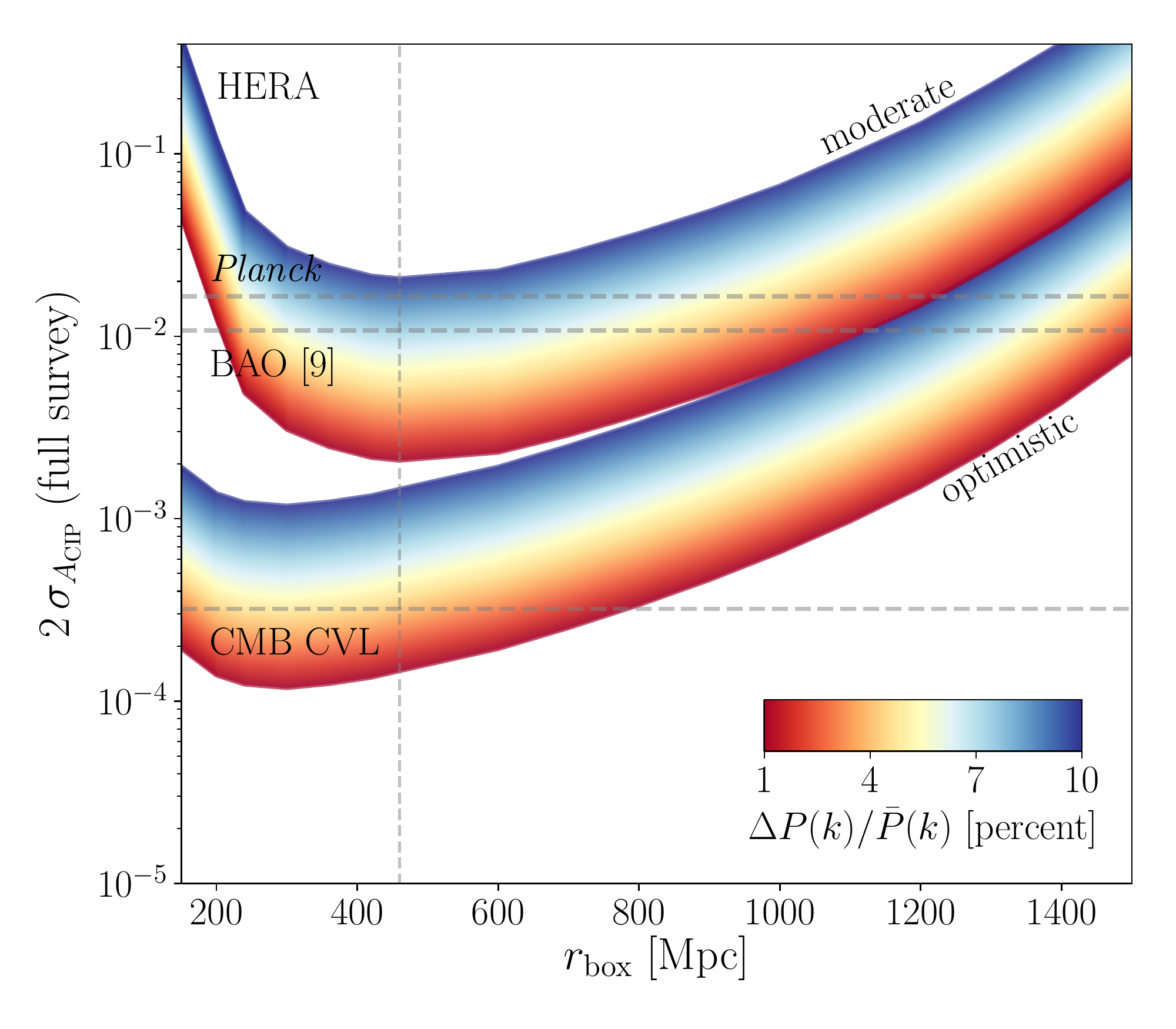}
\end{subfigure}%
\begin{subfigure}{}
  \centering
  \includegraphics[width=.46\linewidth]{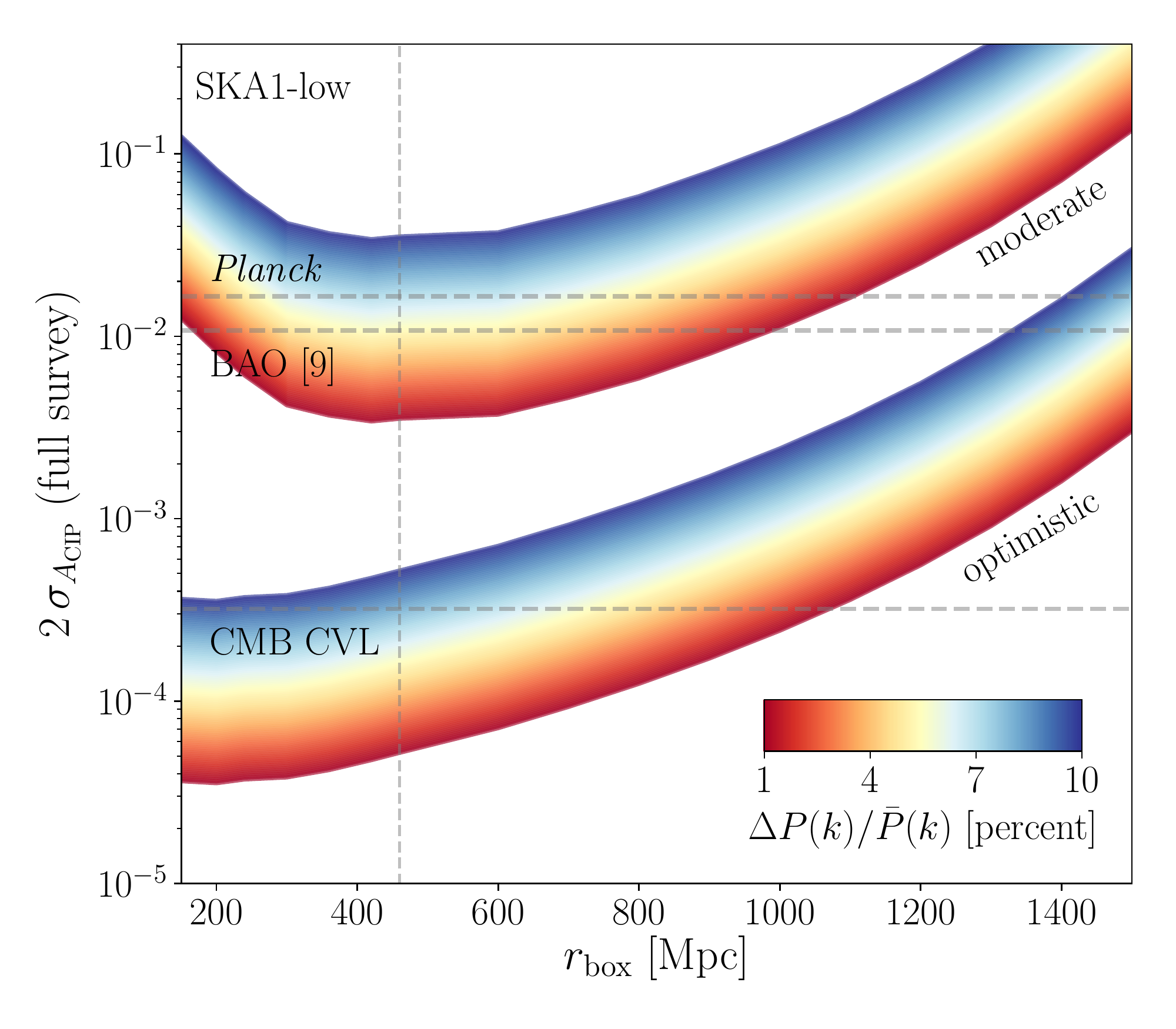}
\end{subfigure}
    \vspace{-0.3cm}
    \caption{ 
    Detection significance for the CIP fluctuations from the HERA (\textit{Left}) and SKA1-low  (\textit{Right}) surveys, shown as a function of the box size in comoving Mpc, using the Fisher matrix formalism defined in Eq.~\eqref{eq:fisherM}. Results are from simulations with medium baryonic-feedback levels and using \texttt{21cmSense}. We find similar constraints for low and high feedback models [constraints improve (worsen) by a less than a factor of $\sim2$ for low (high) feedback] as we marginalised over the smooth spectra. 
    The \textit{Planck} and cosmic-variance-limited (CVL) constraints from the CMB were calculated in~\citep{Smith:2017ndr}, and the BAO constraints (from galaxy surveys) in \citep{Heinrich:2019sxl} (Note that the constraints in Ref.~\citep{Heinrich:2019sxl} are for \textit{correlated} CIP fluctuations). We find SKA1-low may improve upon HERA at the optimistic foreground limit while also being more adversely effected by the baryonic feedback for pessimistic feedback scenarios. For display purposes we omitted plotting constraints from the pessimistic-foreground scenario, which we find $\lesssim\mathcal{O}(10)$ worse than the moderate case.
    }
    \label{fig:display_CIPs_hera_ska}
\end{figure*}

Our reconstruction and forecasts procedure can be summarised as:
\begin{itemize}
    \item We produce separate-Universe {\tt 21cmvFAST} simulation boxes with varying $\Delta(\boldsymbol{X})$ in order to calculate the sensitivity of the 21-cm power spectrum to the local CIP amplitude: $\partial \Delta^2_{21\rm cm}/\partial \Delta(\boldsymbol{X})$.
    \item We calculate the error on the CIP amplitude in each box, $\sigma_\Delta$, using a Fisher-matrix formalism, defined in Eq.~\eqref{eq:fisherM}, marginalising over the non-CIP part of the 21-cm power spectrum with priors we describe in Section~\ref{sec:priors}. 
    \item In order to infer the detection significance of the large-scale CIPs over the full survey volume; we combine measurements of local CIP amplitudes in a patch-work formalism and calculate the reconstruction noise on the large-scale CIPs in Fourier space in Eq.~\eqref{eq:recon_noise}. We model the large-scale CIPs as scale invariant in Eq.~\eqref{eq:CIPsignal}.     
    \item We repeat this procedure for varying box sizes, redshifts, as well as foreground and feedback scenarios. For a given box size and a fixed foreground and feedback scenario, we forecast the total detection significance on the global CIP amplitude, $A_{\rm CIP}$, using 5 redshift bins in the range $z\in[10,27]$. We show the dependence of our final results on the box sizes as well as feedback and foreground scenarios in Figures~\ref{fig:display_CIPs_hera_ska},~\ref{fig:hera_priors}~and~\ref{fig:SNR_ACIP_VAO}.    
\end{itemize}

\begin{table}[b!]
  \begin{center}
    \caption{The upper-limit forecasts ($95\%$ {confidence level}) for the CIP amplitude $A_{\rm CIP}$ as described in the text. 
    We calculate $A_{\rm CIP}$ for various baryonic feedback scenarios and for various foreground levels using the experimental specifications for HERA and SKA1-low described in Appendix \ref{app:telescopes}. 
    The box size is chosen to be $r_{\rm box}=460$Mpc.}
    \label{tab:detection-SNR}
    \begin{tabular}{| c | c c c |} 
    \hline
    HERA &  &  Feedback & \\ 
    \hline
    Foregrounds & ~Low & Medium & High~\\
   \hline
   %1.78, 2.75, 2.29 1e-3
   %2.69, 2.70, 5.14 1e-2
   Pessimistic & $2.7\!\times\!10^{-2}$ & $2.7\!\times\!10^{-2}$ & $5.1\!\times\!10^{-2}$ \\ 
   Moderate & $1.8\!\times\!10^{-3}$ &
   $2.3\!\times\!10^{-3}$ &
   $2.8\!\times\!10^{-3}$  \\ 
   Optimistic & $1.9\!\times\!10^{-4}$ & $2.7\!\times\!10^{-4}$ & $3.4\!\times\!10^{-4}$ \\
   \hline
    \hline
    SKA1-low &  &  Feedback & \\ 
    \hline
    Foregrounds & ~Low & Medium & High~\\
   \hline
   Pessimistic & $2.9\!\times\!10^{-1}$ & $5.6\!\times\!10^{-1}$ & $7.1\!\times\!10^{-1}$ \\ 
   Moderate & $5.5\!\times\!10^{-3}$ & $6.3\!\times\!10^{-3}$ & $9.5\!\times\!10^{-3}$ \\ 
   Optimistic & $6.1\!\times\!10^{-5}$ & $9.5\!\times\!10^{-5}$ & $1.3\!\times\!10^{-4}$ \\ 
   \hline
   \end{tabular}\\ 
  \end{center}
\end{table}

\section{Discussion and Conclusions}\label{sec:discussion}

In this paper we have utilised the constraining power of the 21-cm brightness temperature measurements from the cosmic dawn to evaluate the detection prospects of CIPs. We evaluated the detection significance of CIPs from measurements of both the total change in the power spectra, as well as the shift in the VAO scale, in local boxes of varying size. We have shown that the ongoing HERA and upcoming SKA1-low experiments may be able to measure the uncorrelated CIP amplitude up to a precision comparable to the CMB and LSS experiments.

In this paper we have focused on the effect of CIPs on the acoustic scale, and therefore on the 21-cm signal. 
In practice, however, there can be a multitude of effects that impact the BAO scale locally.
For instance, regions with long-wavelength  (with $k\ll2\pi/r_{\rm drag}$) under- or over-densities of matter can mimic closed or open universes, which could vary the BAO scale and the cosmological parameters locally~(see e.g.~\citep{2012PhRvD..85j3523S}), {which may introduce a bias to our measurement of the power spectra. 
Furthermore,} short-wavelength fluctuations ${k\!\gtrsim\! 2\pi/r_{\rm drag}}$ can also contribute to the noise on the VAO measurement in each box by locally stretching or shrinking the BAO scale, see for a discussion on this matter see e.g.~\citep{Eisenstein:2006nk}. This can potentially contribute to lowering the SNR by boosting the local error on the CIP measurement. 
Further, we have ignored variation in the six cosmological  $\Lambda$CDM parameters, as  we anticipate the priors from the CMB and LSS will dominate the constraints, allowing our assumption of fixed cosmology to be sufficiently robust for our purposes in this paper.

If CIPs are \textit{correlated} with the adiabatic perturbations, they can contribute to the scale dependence of the galaxy bias on large scales. This particular scale dependence is shown to be degenerate with the effect of local non-Gaussianity (parametrised by the amplitude $f_{\rm NL}$) on the galaxy bias in e.g.~Ref.~\citep{Hotinli:2019wdp}). Constraining $f_{\rm NL}$ is one of the main goals of many upcoming large-scale cosmology experiments and forecasts suggest the strongest constraints will be provided from the scale-dependence of the bias. Hence all current and upcoming constraints on $f_{\rm NL}$ can be strongly influenced by the CIPs, which are usually assumed zero. 
Hence, it is important to use a multitude of tracers that may be affected differently when constraining CIPs and $f_{\rm NL}$. Measurements of the 21-cm hydrogen line discussed in this paper can serve to set external priors on both correlated and uncorrelated CIPs improving constraining power of the galaxy and CMB surveys both on CIPs and the primordial non-Gaussianity. 

Upcoming novel observational opportunities will allow significant advances in our understanding of the fundamental properties of the Universe. Among others, characteristics of relative baryon and DM fluctuations prove valuable for probing deviations from adiabaticity that may be sourced by fundamental physics during the early Universe. Constraining CIPs may rule out models of multi-field inflation, or allow less ambiguous measurements of early-Universe signatures such as primordial non-Gaussianity.

The uncorrelated CIPs considered in this paper are difficult to constrain on large scales since the sample-variance cancellation techniques (as done in~\citep{Hotinli:2019wdp}, for example) cannot be utilised in this case to increase the detection significance. Hence, adding to the number of independent measurements is generally valuable. The current constraints on uncorrelated CIP fluctuations afforded by \textit{Planck} and galaxy surveys still allow for the CIP amplitude to be significantly larger than the adiabatic fluctuations we measure in the Universe. 
As discussed in the text, constraints provided by the upcoming 21-cm experiments have the potential to improve current constraints by orders of magnitude in the next decade.\\ 

\section{Acknowledgements}

SCH is supported by the Horizon Fellowship from Johns Hopkins University. SCH also acknowledges the support of Imperial College President's Fellowship, Balzan Fellowship from Oxford University and Johns Hopkins, the Perimeter Visiting Graduate Fellowship and a postdoctoral fellowship from Imperial College London. 
TB acknowledges STFC for their studentship funding. 
JBM was funded by a Clay Fellowship at the Smithsonian Astrophysical Observatory. 
BRD is supported by the Balzan Fellowship from Oxford University. BRD thanks Johns Hopkins University where the project has been started. BRD would like to acknowledge DAE, Govt. of India for financial support through Visiting Fellow through TIFR.
MK was supported by NSF Grant No.\ 1818899 and the Simons Foundation.

\bibliography{vao_constraints.bib}

\appendix

\section{Interferometer noise calculations}\label{app:telescopes}

\subsection{Preliminary SKA1-low noise calculation}\label{sec:ska1low}

The detection of the 21-cm signal (through its power spectrum) with the interferometric optical instruments like SKA~\cite{2019arXiv191212699B,Bacon:2018dui} depends mainly on two types of noise (after foreground removal): system noise and sample variance. These are the main two uncertainties to the foreground substracted 21-cm power spectrum. The system noise is completely instrumental (i.e., it does not depend on the 21 cm signal), however, the sample variance is related to the 21-cm power spectrum. The sample variance dominates on the large scales, whereas the system noise dominates on relatively smaller scales. The expression for the (anisotropic) system noise (denoted by $\delta \tilde{P}_{N}$) is given by \citep{McQuinn:2005hk,2011MNRAS.418..516G,Villaescusa-Navarro:2014cma,Dinda:2018uwm}

\begin{equation}
\delta \tilde{P}_{N}(k,\theta,\nu) = \frac{1}{\sqrt{N_{m}(k,\theta)}} \left( \frac{\lambda^{2}}{A_{e}} \right)^{2} \frac{r_{\nu}^{2} L T_{\rm sys}^{2}}{B t_{0} \tilde{n}(k,\theta,\nu)},
\label{eq:aniso_sys_noise}
\end{equation}

\noindent
where $\nu(z) = \nu_{21}/(1+z)$ is the observed wavelength of the 21 cm signal emitted at redshift $z$. 
The value of the emitted (or the comoving) frequency is given by $\nu_{21}=1420$ MHz, which is the corresponding frequency of the so-called 21 cm ($\lambda_{21}=21$cm) wavelength (emitted or comoving value). Similarly, the observed wavelength is given by $\lambda(z)=\lambda_{21} (1+z)$. $r_{\nu}(z)$ is the comoving distance to the emitted redshift. $L(z)$ is the comoving length of the observation corresponding to the bandwidth, $B$. For the computation of the value of $L(z)$, it can be well approximated as $L(z) \approx \frac{c(1+z)^{2}B}{\nu_{21}H(z)}$ \cite{2011MNRAS.418..516G}, where $c$ is the speed of light in vacuum. In this way, the system noise is (almost approximately) independent of $B$. $A_{e}$ is the effective collecting area of an antenna given by $A_{e}=\epsilon A$, where $A$ is the physical collecting area of an antenna and $\epsilon$ is the efficiency factor. $t_{0}$ is the total observation time. $N_{m}(k,\theta)$ is the total number of independent modes in an annulus of constant $(k,\theta)$ in the range $k$ to $k+\Delta k$ and $\theta$ to $\theta+\Delta \theta$. It is given by $N_{m}(k,\theta)=N_{k}(k) \sin \theta \Delta \theta = 2 \pi k^{2} \Delta k \sin \theta \Delta \theta/V_{1}$, where $V_{1}=(2 \pi)^{3} A_{e}/(r_{\nu}^{2} L \lambda^{2})$ is the resolution in the Fourier space i.e. volume occupied by one independent mode in Fourier space. $T_{\rm sys}$ is the system temperature. $\tilde{n}(k,\theta,\nu)=n_{b}(U=\frac{r_{\nu}}{2\pi} k \sin \theta,\nu)$, where $n_{b}(U,\nu)$ is the baseline distribution. $\vec{U}=\vec{d}/\lambda$ is the baseline vector or the separation vector ($\vec{d}$) (between pair of antennas) in units of wavelength $\lambda$ and $U=|\vec{U}|$. Where circular symmetry has been assumed in $n_{b}(U,\nu)$. In general, baseline distribution can have angular dependency i.e. $n_{b}\equiv n_{b}(\vec{U},\nu)$. $n_{b}(\vec{U},\nu)$ can be rewritten as $n_{b}(\vec{U},\nu)=\frac{N_{t}(N_{t}-1)}{2}\rho_{b}(\vec{U},\nu)$, where $\rho_{b}(\vec{U},\nu)$ is the 2-D baseline distribution related to the antenna distribution and $N_{t}$ is the total number of antenna stations in a specific observation. $\rho_{b}(\vec{U},\nu)$ can be computed as $\rho_{b}(\vec{U},\nu) = B(\nu) \int d^{2}\vec{l}\rho_{\rm ant}(\vec{l})\rho_{\rm ant}(\vec{l}-\lambda \vec{U})$, where $\rho_{\rm ant}(\vec{l})$ is the antenna distributions of the observation. $l$ is the distance from the centre of all the antenna stations. $B(\nu)$ is the redshift dependent integral constant and it is determined by the normalization condition given by $\int d^{2}\vec{U}\rho_{b}(\vec{U},\nu)=1$. 
Eq.~\eqref{eq:aniso_sys_noise} is the anisotropic system noise. 

The isotropic system noise (denoted by $\delta P_{N}$) can be computed by the spherical averaging of Eq.~\eqref{eq:aniso_sys_noise} given by \citep{McQuinn:2005hk,2011MNRAS.418..516G,Villaescusa-Navarro:2014cma,Dinda:2018uwm}

\begin{equation}
\delta P_{N}(k,\nu) = \frac{1}{\sqrt{N_{k}(k)}} \left( \frac{\lambda^{2}}{A_{e}} \right)^{2} \frac{r_{\nu}^{2} L T_{\rm sys}^{2}}{B t_{0} R(k,\nu)},
\label{eq:system_noise}
\end{equation}

\noindent
where $R$ is given by $R(k,\nu)$=$\left[ \sum_{\theta} \sin \theta \Delta \theta \tilde{n}^{2}(k,\theta,\nu) \right]^{1/2}$. Typically, it can be considered that $\sum_{\theta} \sin \theta \Delta \theta \approx \int_{0}^{\pi/2} \sin \theta d \theta$.
The (anisotropic) sample variance (denoted by $\delta \tilde{P}_{SV}$) is given by \citep{McQuinn:2005hk,2011MNRAS.418..516G,Villaescusa-Navarro:2014cma,Dinda:2018uwm}

\begin{equation}
\delta \tilde{P}_{SV}(k,\theta,\nu) = \frac{P_{21}(k,\theta,\nu)}{\sqrt{N_{m}(k,\theta)}}.
\label{eq:aniso_SV}
\end{equation}

\noindent
Similarly, the anisotropic sample variance in Eq.~\eqref{eq:aniso_SV} can be spherically averaged to get isotropic sample variance (denoted by $\delta P_{SV}$) given by \citep{McQuinn:2005hk,2011MNRAS.418..516G,Villaescusa-Navarro:2014cma,Dinda:2018uwm}

\begin{equation}
\delta P_{SV}(k,\nu) \approx \frac{1}{\sqrt{N_{k}(k)}} \left[ \int_{0}^{\frac{\pi}{2}} \frac{\sin \theta d \theta}{P_{21}^{2}(k,\theta,\nu)} \right]^{-\frac{1}{2}}.
\label{eq:sample_variance}
\end{equation}

\noindent
If one neglects the angular dependency in the 21 cm power spectrum (for example, by neglecting the redshift space distortion (RSD) term) or already angle averaging to the 21 cm power spectrum has been done, the isotropic sample variance becomes $\delta P_{SV}(k,\nu)\approx\frac{P_{21}(k,\nu)}{\sqrt{N_{k}(k)}}$. 

The error to the 21 cm power-spectrum measurement is then given by $\delta P_{21}(k,\nu) = \delta P_{SV}(k,\nu) + \delta P_{N}(k,\nu)$.
In our noise calculation for SKA1-low we have considered $\Delta k = k/5$ \citep{Villaescusa-Navarro:2014cma,Sarkar:2015jta,McQuinn:2005hk}.

In this section, we consider the detectability of CIPs through the 21 cm power spectrum with the SKA1-low observations. The SKA1-low specifications are listed below in Table ~\ref{table:tblSKA1low}. The details of the SKA1-low antenna design i.e. SKALA4 (SKA Log-periodic Antenna v4) can be found in \citep{2020arXiv200312512L,Acedo:2020lve}. Typically, we have considered the value of the observation time, $t_{0}$ to be 1000 hours.

% \begin{savenotes}
\begin{table}
\begin{center}
\begin{tabular}{ |p{3cm}|p{3cm}|  }
\hline
\multicolumn{2}{|c|}{SKA1-low instrumental specifications} \\
\hline
Parameter names & Parameter values\\
\hline
Antenna diameter & 40 metre \\
\hline
Physical area of an antenna ($A$) & $\approx 1256.6$ metre$^{2}$ \\
\hline
efficiency factor ($\epsilon$) & $\approx 0.7$ \\
\hline
total no. of antenna stations ($N_{t}$) & 512 \\
\hline
no. of dipole antennas at each station & 256 \\
\hline
total no. of antennas & 131072 \\
\hline
frequency range & 50 MHz - 350 MHz \\
\hline
max redshift range & $\approx$ 3 - 27 \\
\hline
bandwidth ($B$) & 300 MHz$^8$ \\
\hline
\end{tabular}
\end{center}
\vspace*{-0.4cm}
\caption{SKA1-low instrumental specifications.}
\label{table:tblSKA1low}
\end{table}
% \end{savenotes}
\footnotetext{Some authors have used different values of B like 8 MHz in \citep{Liu:2019ygl}, but our calculation is safe because the system noise is almost independent of $B$.}

\noindent
The system temperature ($T_{\rm sys}$) in SKA1-low instrument can be calculated as $T_{\rm sys}(z) = T_{\rm rx}(z) + T_{\rm gal}(z)$, where $T_{\rm rx}(z)=0.1T_{\rm gal}(z)+40K$ is the receiver temperature and $T_{\rm gal}(z)=T_{408}(408 \text{MHz}/\nu(z))^{2.75}$ is the contribution from our own galaxy at a frequency $\nu$ with $T_{408}=25$ K \cite{Bacon:2018dui,Koopmans:2015sua}.

\subsection{\texttt{21cmSense}: Telescope sensitivity estimation} \label{sec:noise_calc}

Foreground mitigation for the cosmological 21-cm signal is performed either via wedge suppression or avoidance\footnote{Please see \cite{2014PhRvD..90b3018L, 2014PhRvD..90b3019L} for detailed description of the EoR window and foreground wedge.}.
\textsc{21cmSense} \citep{Pober:2012zz, Pober:2013jna,  2016ascl.soft09013P} is a python module designed to estimate the noise power spectra when a given telescope array observed the 21-cm signal via foreground avoidance. {We use \textsc{21cmSense} for both HERA and SKA1-low experiments, which we describe below.}

In every u-v bin the noise is calculated as,
\begin{equation}\label{eq: 21cmsense_uv_bins}
\delta^2_{\rm uv}(\mathbf{k}) \approx X^2Y \frac{\mathbf{k}^3}{2 \pi ^2} \frac{\Omega_{\rm Eff}}{2t_{0}} T_{\rm sys}^2,
\end{equation}
where $X^2Y$ is a scalar conversion from an observed solid angle (or effective beam, $\Omega_{\rm Eff}$) to a comoving distance \cite{2014ApJ...788..106P}. 
All other symbols are defined in Section \ref{sec:ska1low}.

Assuming Gaussian errors on cosmic variance, we express the total uncertainty with an inversely weighted sum across all the k modes as
\begin{equation}\label{eq: Tscope_noise}
\delta \Delta_{21}^2(k) = \left\{ \sum_i\frac{1}{[\delta^2_{{\rm uv},i}(k)+\Delta^2_{21}(k)]^2}\right\}^{-\frac{1}{2}} ,
\end{equation}
where the index, $i$, represents multiple measurements of the same frequency from redundant baselines within the array.
This is therefore the total noise, including both sample variance and thermal noise.

\texttt{21cmSense} can implement foreground-wedge avoidance with three levels of severity: 
\begin{itemize}
    \item \textit{Pessimistic} - baselines are added incoherently. No k modes are included from within the horizon wedge (and buffer zone);
    \item \textit{Moderate} - all baselines are added coherently. No k modes are included from the horizon wedge (and buffer zone);
    \item \textit{Optimistic} - All baselines in the primary field of view (no buffer zone) are added coherently.
\end{itemize}

To reiterate, we can write the foreground wedge simply as 
\be\label{eq: wedge}
k_\parallel = a + b k_\perp\,,
\ee
where $k_\parallel$ and $k_\perp$ are the Fourier modes projected on the line-of-side and the transverse plane respectively;  
$b$ depends on the instrument beam, bandwidth and underlying cosmology; 
$a$ is the user defined buffer zone. 
Typically in the \textit{Pessimistic} or \textit{Moderate} case $a = 0.1 h \rm Mpc^{-1}$, 
meaning modes below the line (in Equation \ref{eq: wedge}) are rejected as they are likely contaminated by foregrounds. In this work we consider the moderate scenario with $a=0.03h\rm Mpc^{-1}$ and the \textit{Pessimistic} scenario with $a=0.1h\rm Mpc^{-1}$, while keeping the other parameters involved in defining the noise unchanged.

\begin{figure}[t]
    \vspace{-0.8cm}
    \includegraphics[width =1\columnwidth]{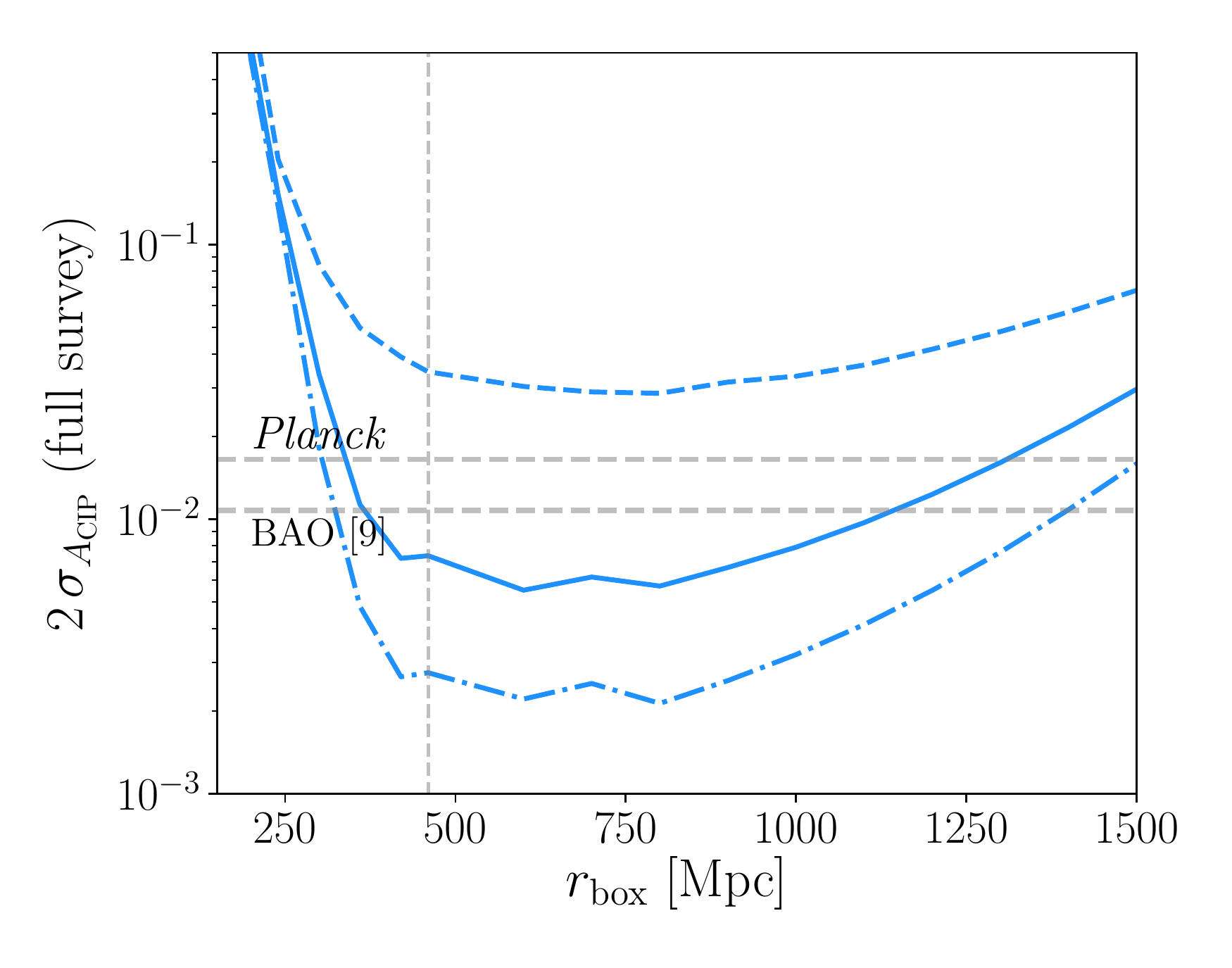}
    \vspace{-1cm}
    \caption{Constraints on CIPs, shown for different assumptions on parameters describing the smooth spectra and the VAO amplitude. The solid line corresponds to the prior choice made in Figure~\ref{fig:display_CIPs_hera_ska} and described in Section~\ref{sec:forcasts}. The dashed line corresponds to no prior assumption on the parameters in the forecast. The dot-dashed line corresponds to the same prior choice as the solid line for the smooth part of the 21-cm spectra, but with $1\%$ prior on the VAO amplitude $A_{\rm vel}(z)$ (an order or magnitude better than for the solid line).} 
    \label{fig:hera_priors}
    \vspace{-0.3cm}
\end{figure}

We apply \textsc{21cmSense} to two telescopes.
Firstly, for SKA we use the specification of SKA1-low where the system temperature is detailed in Section \ref{sec:ska1low}. 
Only the core 225 stations are used as the small baselines generate 21-cm sensitivity for high redshift observations. 
Including longer baselines significantly slows computation and adds negligible precision to the measurement at the redshifts used in this work. 
Each station\footnote{Station locations are taken from  \textcolor{blue}{skatelescope.org/key-documents}} has diameter of 35 $\rm m$ giving SKA a core collecting area of 374444 $\rm m^2$ accross a total bandwidth ranging $[50, 350]~ \rm MHz$. SKA is simulated for 6 hours per night for a tracked scan (different fields for 1 hour each) and as a drift scan.
Secondly, HERA \cite{2016ApJ...826..181D, 2015ApJ...800..128B}, 
where stations are located in a filled hexagonal grid (11 along each side).
Each station is 14 $\rm m$ in diameter giving a total collecting area of 50,953 $\rm m^2$ accross a total bandwidth ranging $[50, 250]~ \rm MHz$.
The antennae are taken to be at $T_{\rm rx} = 100 \rm K$.
HERA is operated only in drift scan mode for 6 hours per night.

\begin{figure*}[t]
  \includegraphics[width=\columnwidth]{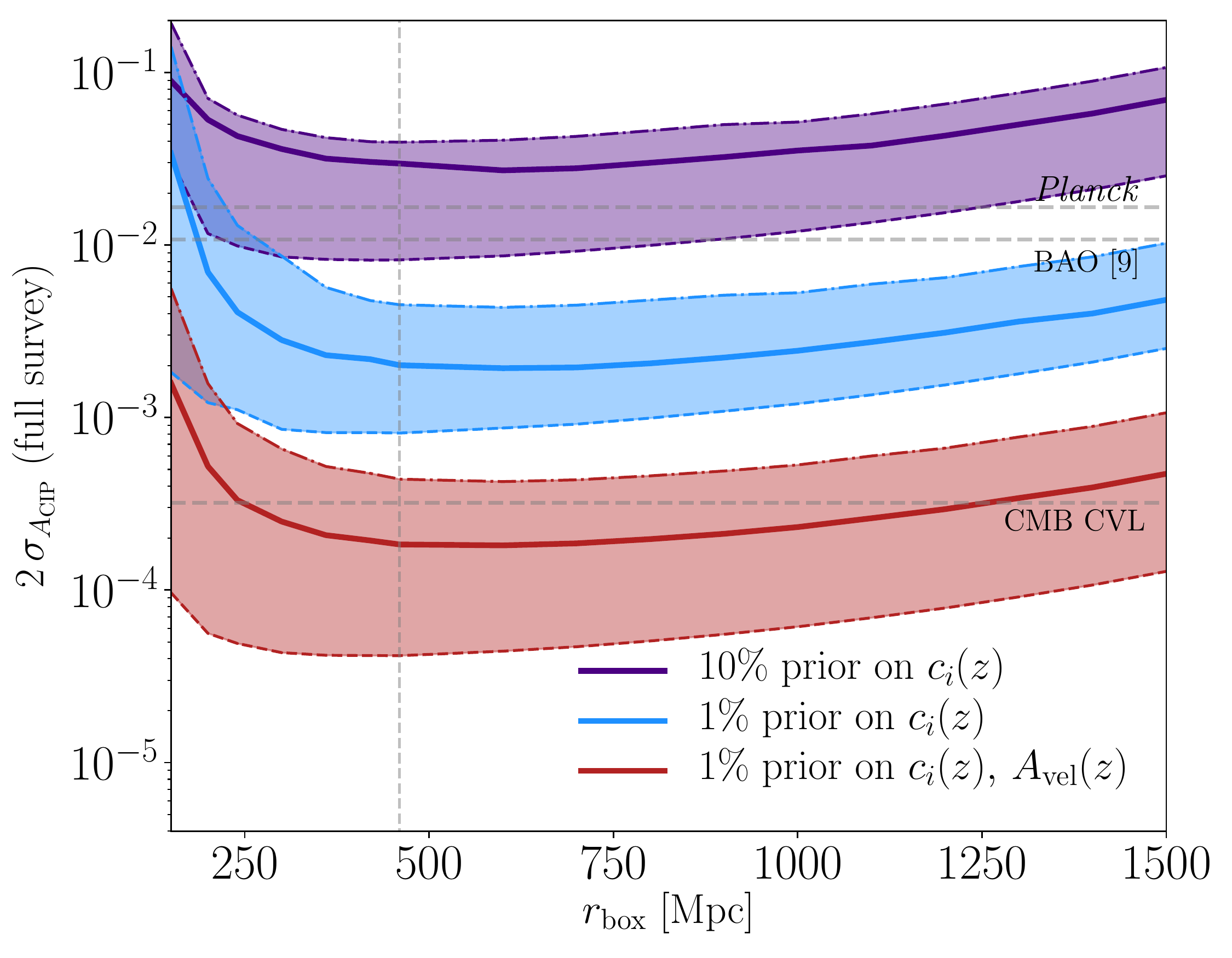}
  \includegraphics[width=\columnwidth]{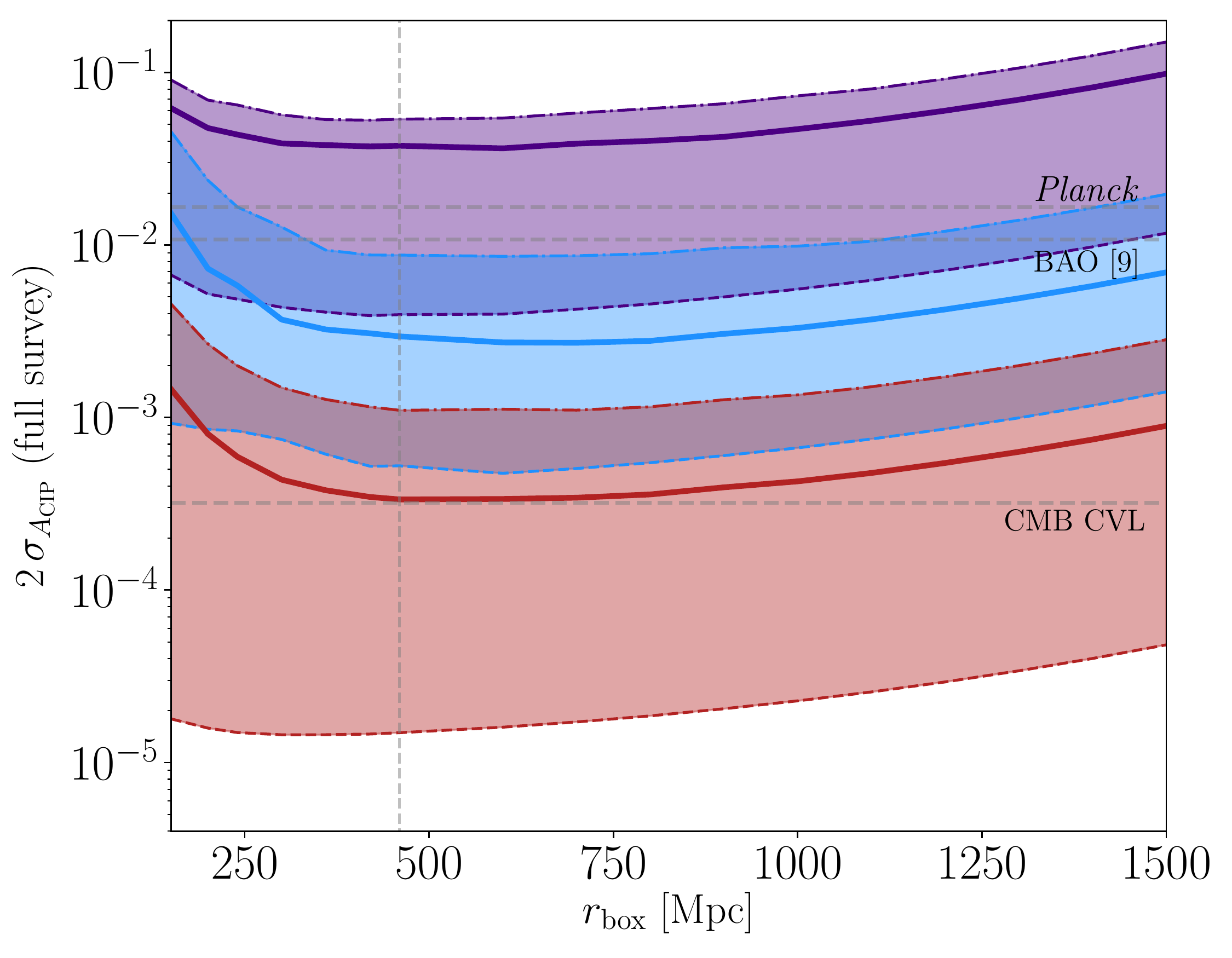}
  \vspace*{-0.4cm}
    \caption{The detection significance for the CIP fluctuations from the ongoing HERA (SKA1-low) survey, shown as a function of the box size in comoving Mpc in the left (right) subfigure. Results are from simulations with medium baryonic-feedback levels using \texttt{21cmSense}. The violet, blue and red shaded regions correspond to three different considerations of priors on our phenomenological model described in Section~\ref{sec:reconstruction}. The violet shaded region corresponds taking 10 percent priors on the parameters describing the smooth component of the spectra, $c_i(z)$ which we assume can be achieved by better and theoretically motivated modelling. The blue shaded region assumes 1 percent prior on the same parameters. Both blue and violet regions do not assume any prior knowledge on the VAO amplitude, $A_{\rm vel}$. The red shaded region corresponds to 1 percent prior on all parameters except the CIP amplitude. The upper dot-dashed lines in each region corresponds to the pessimistic scenario for foreground contamination. The middle solid and lower dashed lines correspond to moderate and optimistic scenarios of foreground contamination.}
    \vspace*{-0.4cm}
    \label{fig:SNR_ACIP_VAO}
\end{figure*}

Both instruments are considered to observe the sky for 3 years and both have the bandwidth taken as 8 $\rm Mhz$ (centred on the frequency relating to the redshift of each coeval simulation box).

\section{Ad-hoc priors}

We demonstrate the role of priors on the CIP constraints in Figures~\ref{fig:hera_priors}~and~\ref{fig:SNR_ACIP_VAO}. We consider a range ad-hoc priors on the non-CIP power-spectrum parameters: i.e. we set priors on $c_i$ by hand rather than physically through $\Delta^2_{21}(k)$ variations, as done in the main text.

Constraints on the smooth spectra and the VAO amplitude can potentially be improved through better modelling the cosmology and reionization. In Figure~\ref{fig:hera_priors}, we show constraints (using the experimental specifications for HERA with moderate foregrounds and medium feedback) from not including any priors (dashed line) along with the prior choice we made in section~\ref{sec:forcasts} (solid line) and a scenario where we added the latter a $1\%$ prior on the VAO amplitude (dot-dashed line). In Figure~\ref{fig:SNR_ACIP_VAO}, we consider two levels of priors: $\{1\%,\,10\%\}$ on all $c_i(z)$ and $10\%$ prior on $A_{\rm vel}(z)$. We forecast for both HERA and SKA1-low experiments, considering a range of box sizes and three levels of foreground contamination: optimistic, moderate and pessimistic, as described in Appendix~\ref{sec:noise_calc}. The blue and purple coloured shaded regions in Figure~\ref{fig:SNR_ACIP_VAO} correspond to CIP constraints taking 1 and 10 percent priors on the parameters $c_i(z)$. The red shaded region corresponds to constraints on CIPs from taking 1 percent priors on all parameters except the CIP amplitude. The upper dot-dashed lines in each shaded region correspond to pessimistic assumptions for foreground contamination, while the middle solid and lower dot-dashed lines correspond to moderate and optimistic foreground contamination assumption, respectively. Overall, we find the modelling of the 21-cm spectra will play a significant role in our ability to isolate the effect of CIPs and the prospects of detecting them.

\end{document}